\documentclass[aps,prl,twocolumn,floatfix,amsmath,amssymb, showpacs,tightenlines]{revtex4}
\usepackage{mathrsfs}
\usepackage{bbm}
\usepackage{amssymb}
\usepackage{graphicx,color}
\usepackage{epsfig,graphicx,times}
\usepackage{amstext}
\usepackage{amsmath}
\usepackage{latexsym}
\usepackage{bm}
\usepackage{CJK}

\begin{document}

\begin{CJK*}{GBK}{song}
\title{A Novel Hohlraum with Ultrathin Depleted-Uranium-Nitride Coating Layer for Low Hard X-ray Emission and High Radiation Temperature}
\author{Liang Guo,$^1$ Yongkun Ding,$^1$$^,$$^*$ Peifeng Xing,$^1$ Sanwei Li,$^1$ Longyu Kuang,$^1$ Zhichao Li,$^1$ Taimin Yi,$^1$ Guoli Ren,$^2$ Zheqing Wu,$^2$ Longfei Jing,$^1$ Wenhai Zhang,$^1$ Xiayu Zhan,$^1$ Dong Yang,$^1$ Bobin Jiang,$^1$ Jiamin Yang,$^1$ Shenye Liu,$^1$ Shaoen Jiang,$^1$ Yongsheng Li,$^2$ Jie Liu,$^2$ Wenyi
Huo,$^2$ Ke Lan,$^2$ Weiyan Zhang,$^2$ and Xiantu He$^2$}

\address{$^1$Research Center of Laser Fusion, Chinese Academy of Engineering
Physics, Mianyang 621900, China\\
$^2$Institute of Applied Physics and Computational Mathematics, Beijing, 100088,
China}

\begin{abstract}

An ultrathin layer of uranium nitrides (UN) has been coated on the inner surface of the depleted uranium hohlraum (DUH), which has been proved by our experiment to prevent the oxidization of Uranium (U) effectively. Comparative experiments between the novel depleted uranium hohlraum and pure golden (Au) hohlraum are implemented on SGIII-prototype laser facility. Under the laser intensity of 6$\times$10$^{14}$ W/cm$^2$, we observe that, the hard x-ray (\emph{h}$\nu$ $>1.8$ keV) fraction of this uranium hohlraum decreases by 61\% and the peak intensity of total x-ray flux (0.1 keV$\sim$5.0 keV) increases by 5\%. Radiation hydrodynamic code LARED is used to interpret the above observations. Our result for the first time indicates the advantage of the UN-coated DUH in generating the uniform  x-ray field with a quasi Planckian spectrum and thus has important implications in optimizing the ignition hohlraum design.

\end{abstract}


\maketitle

\end{CJK*}
\emph{Introduction}---In the indirect-drive inertial confinement fusion (ICF) scheme, intense laser beams first heat the inner wall of a high-Z hohlraum to emit x rays, and subsequently the x rays ablate and accelerate the plastic (CH) shell of a Deuterium-Tritium (DT) capsule placed in the center of the hohlraum. The shell kinetic energy is then converted into the inner energy of DT fuel through an adiabatic compression process that lifts the fuel temperature up to a few keV and fuel density to a few hundred times solid density, and finally triggering DT nuclear fusion and leading to capsule implosion \cite{nuckolls,lindl2004}. In the above process, a homogeneous x-ray field with a Planckian spectral distribution is highly desired, which however is always altered by $M$-band emission in traditional Au hohlraums \cite{kauffman1994,huowy2012}. The $M$-band x rays usually take around 10\%$\sim$20\% fraction of the total x rays in Au hohlraum expriments \cite{lindl2014,moody2014}. The corresponding spectrum can be considered as a superposition of a Planckian spectrum and a harder x-ray Gaussian spectrum. It has been recognized that hard x rays (\emph{h}$\nu$ $> $1.8keV) can penetrate the capsule ablator and preheat DT fuel prior to the arrival of ignition shock, causing the fuel areal density to drop seriously \cite{olsen2003,gu2012}. To prevent the preheating effect, multi-layer doped ablators have been applied in ignition capsules, such as copper-doped beryllium (Be), germanium-doped CH and silicon-doped CH \cite{hurricane2014,park2014,dittrich2014,lill2014}. Nevertheless, adding dopant is a challenging technique that brings in uncertainties. In some cases, doping layers might lead to serious hydrodynamic instabilities at the interfaces between dopants and Be/CH in the ablator and enhance undesired mixing of the fuel and the ablator materials. In addition, the $M$-band x rays are mainly generated in the coronal plasma near the laser focal spots \cite{zhaoyq2014}, therefore the intensity of $M$-band x rays is strongly anisotropic in space and it may arouse serious low-mode implosion asymmetry.

Depleted uranium is expected to be a better choice than Au among many other candidate materials to optimize the x-ray filed. This is because depleted uranium has higher albedo and lower hard x-ray emission under the radiation temperature higher than 170 eV \cite{wilkens2007,dewald2008,lix2010}. Nevertheless, depleted uranium is likely to be oxidized and the uranium oxides are chemically unstable. The oxidization process will induce internal stress in the bulk uranium, which gives rise to the occurrence of cracks, crinkles and delamination of the hohlraum interior. To hurdle the problem, the ignition hohlraums recently applied in National Ignition Campaign (NIC) are of a sandwiched geometry of Au+U+Au \cite{lix2010}. The inner Au coating is usually 600 nm$\sim$700 nm thick, designed to protect the middle U layer against oxidization \cite{lindl2014,kline2013}. However, the Au coating layer with a thickness of several hundred nanometers can emit intense $M$-band x rays, counteracting the advantage of middle U layer in hard x-ray emission. As a result, the hard x-ray fraction of the Au+U+Au hohlraum exceeds 14\% when the radiation temperature is about 300 eV \cite{lindl2014}.

In this Letter, we report a novel depleted uranium hohlraum (DUH), which preserves the sandwiched structure while replaces the inner Au coating by one ultrathin layer of depleted-uranium-nitride (UN). We have proved experimentally  the UN can prevent the oxidization of DUH effectively. Under the laser intensity of 6$\times$10$^{14}$ W/cm$^2$, we observe that, the hard x-ray fraction above 1.8 keV of this uranium hohlraum decreases by 61\% and the peak intensity of total x-ray flux (0.1 keV$\sim$5.0 keV) increases by 5\%. Our two-dimensional radiation hydrodynamic code LARED is used to interpret the above observations. We also extend our discussions to ignition hohlraum.

\emph{UN-coated DUH against oxidation}---We choose uranium nitrides as the coating material because UN has the potential to prevent uranium oxidation according to recently reported work \cite{liu2013}. We coat a UN layer of 100 nm thickness on the inner surface of a cylindrical depleted uranium hohlraum. The fluctuation of the coating thickness is controlled below 10 nm. Each newly fabricated UN-coated DUH is exposed to laboratory air to test their oxidation endurance. After 72 hours, the wall composition of a UN-coated DUH is diagnosed by an x-ray photoelectron spectroscopy (XPS) \cite{wielicaka1995}. Figure 1 shows the atomic percentages of Oxygen, Nitrogen and Uranium in the hohlraum material change with the distance from the inner wall surface. The oxygen content drops rapidly from 56.5\% at the surface to 6.7\% at about 3 nm below the surface, which is due to the free-state oxygens adsorbed on the UN coating and the oxides of UN surface. From the depth of 3 nm to the depth of 75 nm, the molar ratio of Uranium and Nitrogen is around 1:1, while the average oxygen content is kept below 1\%, verifying the effective prevention of the oxygen diffusion and the uranium oxidation by UN coating. Within the entire duration of our comparative experiment, no cracks or delamination are observed on the inner wall of the novel uranium hohlraum under the complicated working conditions including mechanical vibration. The above results demonstrate the UN-coated DUH is fairly reliable.

\begin{figure}[t]
\includegraphics[width=8.5cm]{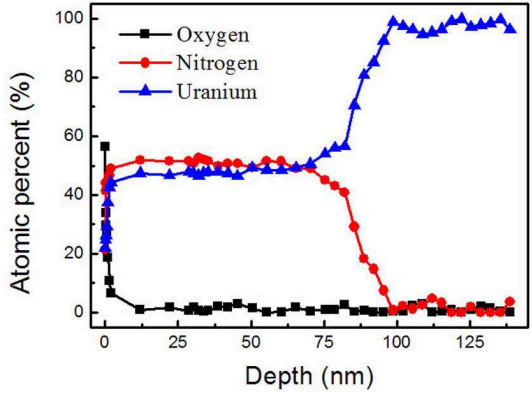}
\caption{(color online). The atomic percentage evolutions with the depth from inner wall surface for the major elements of UN-coated DUH acquired by XPS.}
\label{setup}
\end{figure}

\emph{Experimental setup}---The Rosseland mean opacity of UN is analogous to that of Uranium, and Nitrogen has almost no hard x-ray emission. We thus expect the influence of UN layer on the x-ray field produced by DUH should be much smaller than that of Au layer with the same thickness. Therefore, the UN-coated DUH is able to produce an x-ray field with lower hard x-ray fraction and higher radiation temperature. To check the above idea, we carry out the comparative experiments on SGIII-prototype laser facility using both UN-coated DUH and pure Au hohlraum. Figure 2 shows the experimental setup. The diameter and length of the empty hohlraum are both 1000 $\mu$m, and the diameter of the laser entrance hole is 800 $\mu$m. The uranium hohlraum wall consists of three layers: 100 nm-thick UN as the coating layer, 2 $\mu$m-thick depleted uranium in the middle and 25 $\mu$m-thick Au as the outer layer. Eight 351 nm laser beams inject into the hohlraum with an incident angle of $45^{\circ}$ with respect to the hohlraum axis. All the lasers are square wave with a duration of 1 ns, and the total laser energy is 4.9 kJ. The diameter of the laser focal spot is adjusted to 500 $\mu$m by using the continuous phase plates. As shown in Fig. 2(a), the laser spots are initially located in the middle of the hohlraum with the neighboring two overlapped to form a closed loop. The laser intensity in the overlapping area reaches 6$\times$10$^{14}$ W/cm$^2$. A transmission grating spectrometer (TGS) is installed at $20^{\circ}$ to acquire the spectrum from 0.2 keV to 5.0 keV \cite{weaver1995}. The TGS consists of one calibrated 2000 l/mm transmission grating and one detecting charge-coupled device (CCD). The spatial and spectral resolution of TGS are 110 $\mu$m and 0.075 nm, respectively. Three groups of x-ray detectors (XRD) are set at $20^{\circ}$,  $30^{\circ}$ and  $45^{\circ}$. Each group comprises two types of detectors. Flat response XRD (FXRD) measures the intensity of the total x-ray flux from 0.1 keV to 5.0 keV, through which the radiation temperature inside hohlraums can be obtained \cite{lizc2010}. $M$-band flat response XRD (MXRD) measures the intensity of $M$-band x-ray flux from 1.6 keV to 4.4 keV \cite{guol2012}. An artificial spectrum comprised of a Planckian distribution and a Gaussian distribution is involved to calculate the intensity of x-ray flux above 1.8 keV. The time resolution and intensity uncertainty of XRD is 100 ps and 10\%, respectively. Two x-ray pinhole cameras (XPHC) are employed to monitor the spatial accuracy of the lasers.
\begin{figure}[t]
\includegraphics[width=8.5cm]{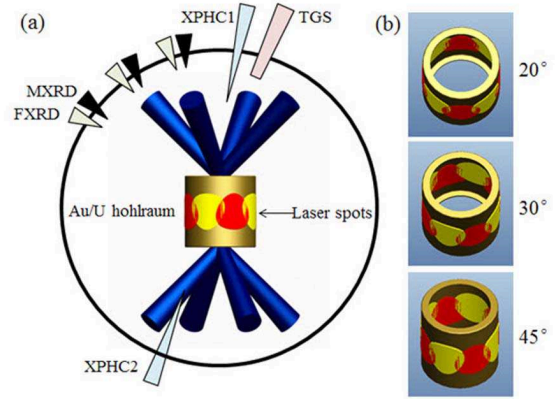}
\caption{(color online). Schematic of experimental setup (a) and view fields of detectors (b).The red and yellow spots are produced by the top and bottom laser beams, respectively.}
\label{shock}
\end{figure}

\emph{Results and analysis}---Figure 3 shows the radiation temperatures ($T_r$) measured by FXRD and simulated by LARED (two dimensional radiation hydrodynamic code \cite{yangh2013}). According to our previous measurements, the backscattered and near-backscattered energy is set as 8\% of the total laser energy in the simulation. The measured peak $T_r$ of Au hohlraum by detectors set at $20^{\circ}$, $30^{\circ}$, and $45^{\circ}$, is 204 eV, 218 eV, 214 eV respectively, while that of UN-coated DUH is 207 eV, 220 eV and 218 eV correspondingly. The increased temperature demonstrates UN-coated depleted uranium has slightly higher x-ray conversion efficiency than Au. The simulated $T_r$ agrees well with the experimental results in term of both temporal behavior and angular distribution. The faster rise of simulated $T_r$ than measured $T_r$ after 0.8 ns could be caused by the x-ray emission of the U/Au gas accumulated near the hohlraum axis, which is filled in hohlraum in the simulation with an initial density of 0.008 g/cm$^3$ and does not truly exist in experiments. The $T_r$ of $20^{\circ}$ is lower than that at other two angles, due to the smallest fraction of the laser focal spot area in the view field area at $20^{\circ}$ as shown in Fig. 2(b).
\begin{figure}[t]
\includegraphics[width=8.5 cm]{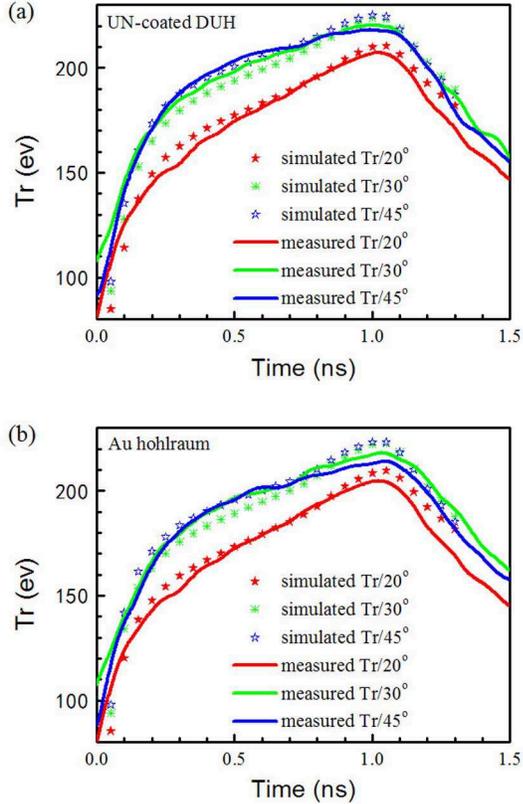}
\caption{(color online). Experimental (solid lines) and simulated (stars) radiation temperatures of UN-coated DUH (a) and Au hohlraum (b). }
\label{xrayinput}
\end{figure}

Figure 4(a) is the time-integrated x-ray images captured by the CCD of the transmission grating spectrometer, showing that the hard x rays are significantly reduced in UN-coated DUH. The view field of the 900$\sim$950 pixels of the CCD can observe the laser spot area. Compared with the Au hohlraum that gives about 30000 counts around 2.5 keV, the UN-coated DUH exhibits almost no hard x-ray emission in the image and the CCD counts are only 3000. Meanwhile, the soft x-ray emission from UN-coated DUH has a distinct enhancement, which is consistent with the increasing trend of the radiation temperature. Figure 4(b) is the unfolded spectrums from Fig. 4(a), which shows $N$-band x-ray emission of UN-coated DUH blue-shifts from 0.8 keV to 1.2 keV and the spectral shape of hard x rays of UN-coated DUH is similar to that of a Planckian distribution. The integrated value of x rays above 1.8 keV from UN-coated DUH is 65\% lower than that from Au hohlraum.
\begin{figure}[b]
\includegraphics[width=8.5cm]{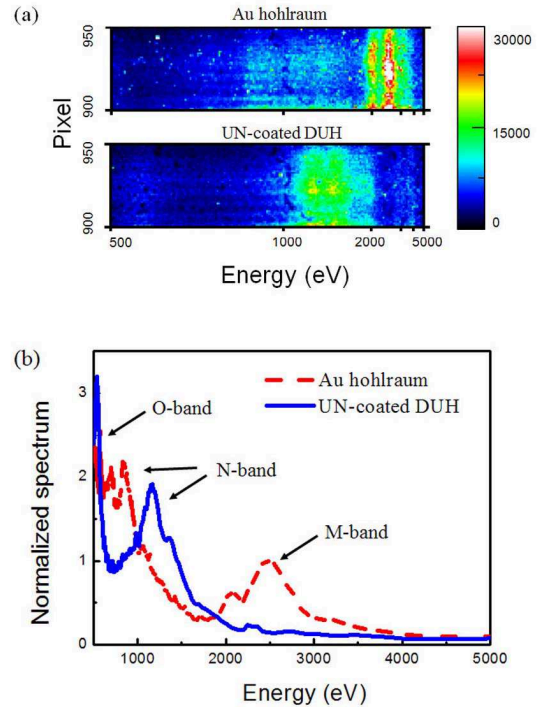}
\caption{(color online). X-ray images captured by CCD of TGS (a), and spectrums of UN-coated DUH and Au hohlraum derived from the CCD images (b) normalized by the peak value of $M$-band spectrum of Au hohlraum.}
\label{trfmplane}
\end{figure}

Figure 5 is the temporal behavior of the spectral-integrated x-ray flux measured by the XRD group at $20^{\circ}$. It is found that the UN-coated DUH produces 5\% higher peak intensity of total x-ray flux than Au hohlraum. And the integrated value of the hard x-ray flux of UN-coated DUH is 54\% lower than that of Au hohlraum. The descending range of 54\% by XRD is smaller than the 64\% by TGS, because XRD determines an average intensity in the entire view field whereas the detection area of TGS is only a small part of the XRD view field area. Hard x-ray fraction is 12\% for Au hohlraum and 4.7\% for UN-coated DUH, which is calculated from the ratio between the integrated values of hard x rays and total x rays. Apparently, the hard x-ray fraction of UN-coated DUH decreases by 61\%. In a Planckian radiation source with a $T_r$ of 207eV, the hard x-ray fraction is 2.5\%, which indicates the hard x-ray emission of UN-coated DUH is very close to the situation of a Planckian source. It is observed that the rising and falling rates of the total flux from UN-coated DUH are faster than those from Au hohlraum. A possible explanation is that the atomic processes of Uranium are faster than those of Au, as the uranium atom has more outer electrons and the minimum ionization energy of uranium atom is relatively smaller. The faster response of laser pulse to x rays of UN-coated DUH is benefit of the shock shaping in ICF.
\begin{figure}[t]
\includegraphics[width=8.5cm]{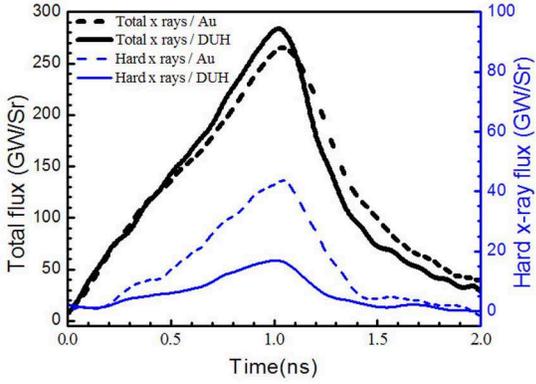}
\caption{(color online). Total (0.1 keV$\sim$5.0 keV) and hard x-ray (\emph{h}$\nu$ $>1.8$ keV) fluxes of UN-coated DUH (solid line) and Au hohlraum (dash line). The black thick curves correspond the left vertical axis and the blue thin curves correspond the right one.}
\label{trfmlaser}
\end{figure}

The above experimental results have demonstrated UN-coated depleted uranium hohlraum can produce an x-ray field with ultra-low hard x-ray fraction and high radiation temperature, under the laser intensity of SGIII-prototype laser facility. We further use DCA code with non-local thermodynamic equilibrium model \cite{wuzq2003} to calculate the hard x-ray emissivities of UN-coated depleted uranium and Au under the ignition laser condition. The simulation shows both the peak and spectral-integrated hard x-ray emissivities of UN-coated uranium are much lower than those of Au, when the electron temperature is up to 3.5 keV and the plasma density is 0.02 g/cm$^3$, as shown in Fig 6. The result indicates the advantage of UN-coated DUH in low hard x-ray fraction can extend to the ignition status. The significantly decreased hard x rays can weaken the preheating effect and reduce the requirement of adding dopants in ignition capsules. In addition, the lower hard x-ray fraction means the more homogeneous intensity distribution of x rays in space, which is benefit to improve the implosion symmetry. We use IRAD3D code \cite{hangyb2014} to interpret the influence on the asymmetry of the capsule by decreased hard x rays. This code is based on the analysis of view factor, in which hard x rays only distribute in the laser focal spot areas while soft x rays distribute in both spot areas and re-emission areas with a weight ratio of 2:1 according to the previous observation \cite{zhaoyq2014}. In a simple case that eight laser beams inject into a cylindrical hohlraum to form two loops of laser spots, the calculated P2 asymmetry of the capsule by x rays drops from 7.27\% to 3.15\%, if the hard x-ray fraction changes from 12\% to 4.7\%. Based on the benefits about preheating and asymmetry, we believe UN-coated depleted uranium is a promising material of the ignition hohlruam. Therefore, we further apply UN-coated depleted uranium to design a new ignition hohlraum. According to the method in our previous work \cite{lix2010}, the wall composition of this hohlraum is 100 nm UN + 5 $\mu$m U + 19.9 $\mu$m Au.  Comparing with the sandwich hohlruam of Au+U+Au applied in NIC, the UN+U+Au sandwich hohlraum can generate nearly the same radiation temperature but lower hard x-ray fraction.

\begin{figure}[t]
\includegraphics[width=8.5cm]{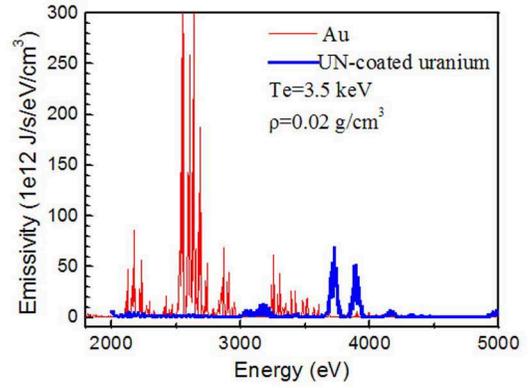}
\caption{(color online). X-ray emissivities of UN-coated uranium (thick blue) and Au (thin red) simulated by DCA code.}\label{trfmplane}
\end{figure}

In summary, we propose that the uranium nitrides as a coating layer can effectively prevent the oxidization of depleted uranium hohlraum and demonstrate the spectral optimization of x-ray source by UN-coated DUH for the first time. We envision that the novel hohlraum is a better choice than pure Au hohlraum or even Au+U+Au sandwich hohlraum for ignition target design. Further work on fabricating gas-filled uranium hohlraums coated by UN and studying the improvements of implosion on the nearly completed SGIII laser facility with total laser energy of about 100 kJ in shaped pulses are currently undergoing.

This work was supported by the National Natural Science Foundation of China under Grant No 11475154.
\begin{acknowledgements}

\end{acknowledgements}

\end{document}